\documentclass[%
 aps,
pra,
superscriptaddress,
 amsmath,amssymb,
 reprint,%
]{revtex4-2}
\usepackage{graphicx} 
\usepackage{amsmath}
\usepackage{dcolumn}
\usepackage{bigints}
\usepackage{float}
\usepackage{color}
\usepackage{etoolbox}
\usepackage{braket}

\usepackage[utf8]{inputenc}
\usepackage[T1]{fontenc}

\usepackage[english]{babel}

\makeatletter
\def\@email#1#2{%
 \endgroup
 \patchcmd{\titleblock@produce}
  {\frontmatter@RRAPformat}
  {\frontmatter@RRAPformat{\produce@RRAP{*#1\href{mailto:#2}{#2}}}\frontmatter@RRAPformat}
  {}{}
}%
\makeatother
\begin{document}

\preprint{AIP/123-QED}

\title[]{
Effects of Screening and Pressure Ionization on the Electron Broadening of Spectral Lines in Dense Plasmas
\\}

\author{J.P. Kinney}
 \email{julkin@umich.edu}
\affiliation{ 
Department of Nuclear Engineering and Radiological Sciences, University of Michigan, Ann Arbor, MI 48109, USA}
\author{S.B. Hansen}%
\affiliation{ 
Sandia National Laboratories, Albuquerque, NM 87185, USA}%

\author{T.A. Gomez}
\affiliation{ 
Department of Nuclear Engineering and Radiological Sciences, University of Michigan, Ann Arbor, MI 48109, USA}
\author{S.D. Baalrud}
\affiliation{ 
Department of Nuclear Engineering and Radiological Sciences, University of Michigan, Ann Arbor, MI 48109, USA}

\date{\today}

\begin{abstract}
Collisions between electrons and radiating atoms broaden spectral absorption and emission lines in dense plasmas. High densities also introduce screening and pressure ionization effects that distort the wavefunctions of both bound and free electrons. In order to study how 
dense plasma effects influence the electron broadening of spectral lines, this paper incorporates 
electron wavefunctions from an average-atom (AA) model to calculate the line width of the B~{\sc iii} $2p-2s$ transition at $T = 10$~eV for mass densities ranging from $\rho=10^{-4}-0.4$~g/cc. The calculation method uses the impact approximation, allowing the line width to be written in terms of electron-collision cross sections and an interference term. Compared to an otherwise identical calculation that uses Coulomb free wavefunctions, the AA method is found to modify both the cross sections and the resulting line width at sufficiently high density by introducing screening and pressure ionized bound states. Screening lowers the cross sections at low energies and near electron excitation thresholds, while pressure ionized bound states introduce resonances into the continuum. Thus, as the density increases, the relative line width between the AA and Coulomb calculations follows a general decrease because of screening, with sharp increases at various intervals due to pressure ionization. The AA results are also compared with a common approach to introduce screening through the interaction potential and reduced models that use the Bethe formula for the inelastic electron-collision cross sections.

\end{abstract}

\maketitle

\section{\label{sec:Intro}Introduction}

Accurate calculations of spectral line shapes are important for both computational modeling and experimental diagnosis of plasmas. In computation, opacities are crucial for modeling radiation transport~\cite{rogers1992radiative,colgan2016lanl,hirose2022optab,Drake2018}. In experiments, comparison between measured and theoretically calculated radiation spectra is often used to diagnose basic plasma conditions such as temperature and density~\cite{yaakobi1977direct,florida2008analysis,falcon2015lab}. Ultimately, a full description of spectral line shapes in plasmas is a very complicated $N$-body problem. Accordingly, much work on line shapes has been devoted to developing and improving simulation-based~\cite{gigosos1986stark,stambulchik2006astudy,stambulchik2007correlation,gomez2016xenomorph,gigosos2018classical} and semi-analytic~\cite{griem1959Stark,BarangerI,BarangerII,BarangerIII,fano1963pressure,hussey1975kinetic,iglesias1985low,boercker1987radiative,talin1995frequency,gomez2021allorder} descriptions. Although simulation work has been performed considering the exact $N$-body dynamics of classical charged particles in a plasma~\cite{stambulchik2007correlation,gigosos2018classical}, many semi-analytic line shape calculations instead approximate the $N$-body dynamics by using an exponential screened potential in a two-body interaction model, and assume a large timescale separation between electron and ion motion~\cite{griem1959Stark,gomez2021allorder,gomez2022intro}. 
This work advances these semi-analytic approaches by modifying electron wavefunctions using an average-atom (AA) model and studying its effect on calculations of the spectral line width in the impact approximation.

The impact approximation presents a widely used theoretical framework to calculate the shift and width of spectral lines due to collisions with plasma electrons~\cite{BarangerI,BarangerII,BarangerIII}. It assumes that collisions between individual plasma electrons and the radiating atom are statistically independent and instantaneous with respect to the radiative transition timescale. The spectral line width is then calculated by considering independent perturbations to the upper and lower atomic states and a correlation correction known as the interference term. An important insight by Baranger~\cite{BarangerIII} is that the upper and lower state perturbations to a spectral line in this framework can be related to electron-collision cross sections through the optical theorem. Importantly, free-electron wavefunctions are needed to calculate both the cross sections and the resulting line width.

The main purpose of this paper is to evaluate the impact approximation using free-electron wavefunctions calculated from a self-consistent average-atom (AA) model that natively includes density effects such as plasma screening and pressure ionization. Average-atom models have been widely applied to calculate radiative transport properties in plasmas~\cite{johnson2006optical,shaffer2017free,tacu2025electrical,hansen2024} and are based on density functional theory, where the properties of a quantum mechanical system are written in terms of the electron density. The AA model used in this paper provides a self-consistent potential around a B~{\sc iii} atom situated in a plasma at $T = 10$~eV for mass densities ranging from $\rho=10^{-4}-0.4$~g/cc. Then, free-electron wavefunctions calculated in this self-consistent potential are used to study the electron-collision cross sections and spectral line width of the isolated B~{\sc iii} $2p-2s$ transition. This transition is largely unaffected by ion-Stark effects and its width is dominated by electron broadening.

Two physical effects of including the AA model are observed. First, as the density increases, the screening in the AA potential 
modifies the electron-collision cross sections at low energies and near electron excitation thresholds. Compared to identical calculations using Coulomb free wavefunctions, this causes the relative line width to generally decrease with increasing density. This conclusion is shown to qualitatively agree with traditional methods of including a screened electron-atom interaction operator in the calculation of matrix elements~\cite{gomez2022intro}. However, it is also shown that including the screening in the interaction operator reduces the line width much more than including screening in the free wave functions themselves. Second, the AA model includes bound states that have been pressure ionized into the continuum. As the density increases, these pressure-ionized bound states show up as resonances in the cross sections that result in sharp increases in the line width.

The remainder of this paper is organized as follows. Section~\ref{sec:Theory} introduces the impact approximation framework, the AA model, the description of pressure-ionized bound states, and the Bethe formula for inelastic electron-collision cross sections. Section~\ref{sec:Discussion} discusses the effect that free electron wavefunctions from the AA model have on calculations of the electron-collision cross sections and the line width. Section~\ref{sec:Discussion} also compares the impact approximation with cross sections and line widths calculated using the Bethe formula. Section~\ref{sec:conclusion} discusses conclusions based on these results. Finally, all equations in this paper use atomic units ($\hbar=m_{e}=e=4\pi\epsilon_{0}=1$).

\section{\label{sec:Theory}Theory}
The calculations of the B~{\sc iii} $2p-2s$ transition involve a central nucleus ($Z = 5$) surrounded by two $1s$ electrons that form a `frozen core' (i.e., assumed to stay in their respective quantum states during a collision with a perturbing free electron) and one radiating electron that transitions between the $2p$ and $2s$ states. In the forthcoming theory, $\alpha,
\beta$ are used to denote the upper and lower states of the atom ($1s^{2}2p$, $1s^{2}2s$). The impact approximation provides the framework for calculating the line width of this transition due to collisions between the atom and free electrons. The AA model is used to determine the quantum states of all the electrons.
        
\subsection{\label{subsec:LineBroadening}Electron Broadened Line Width in the Impact Approximation}

The impact approximation makes two fundamental physical assumptions about collisions between plasma electrons and a radiating atom~\cite{BarangerII}. First, interactions between individual plasma electrons and the atom are taken to be statistically independent. Thus, the physics of the full $N$-body interaction is reduced to a two-body interaction where the potential may include the mean field of the other particles. Second, electron-atom collisions are taken to be instantaneous. This implies that the collision timescale is faster than the timescale of the radiative transition itself. 
In addition to the impact approximation, the calculations here take the B-III $2p-2s$ transition to be an isolated line, unaffected by quasi-static fields from other plasma particles. Then the line shape is Lorenztian
\begin{eqnarray}
    \label{eq:Lorenztian}
    I(\omega)\propto \frac{1}{\left(\omega-\omega_{\alpha\beta}-s\right)^{2}+w^{2}}
\end{eqnarray}
where $\omega_{\alpha\beta} = E_{\alpha}-E_{\beta}$ is the line center given by the energy difference, and $s$ and $w$ are the shift and width of the line. The line width is given in terms of $T$-matrices as~\cite{BarangerIII}
    \begin{flalign}
        \label{eq:width2}
        \nonumber w &= n_{e}\lambda_{T}^{3}\int \frac{d^{3}\mathbf{k}}{\left(2\pi\right)^{3}}e^{-\frac{1}{2}\beta k^{2}}\left[\mathrm{Im}\left\{\bra{\alpha\mathbf{k}}\hat{T}\left(E_{\alpha}+E_{\mathbf{k}}\right)\ket{\alpha\mathbf{k}}\right\}\right.\\&\nonumber-\mathrm{Im}\left\{\bra{\beta\mathbf{k}}\hat{T}^{*}\left(E_{\beta}+E_{\mathbf{k}}\right)\ket{\beta\mathbf{k}}\right\}\\
        \nonumber&+2\pi\int \frac{d^{3}\mathbf{k}'}{(2\pi)^{3}}\bra{\alpha\mathbf{k}}\hat{T}\left(E_{\alpha}+E_{\mathbf{k}}\right)\ket{\alpha\mathbf{k}'}\\&\left.\times\bra{\beta\mathbf{k}'}\hat{T}^{*}\left(E_{\beta}+E_{\mathbf{k}}\right)\ket{\beta\mathbf{k}}\delta\left(E_{\mathbf{k}}-E_{\mathbf{k}'}\right)\right]
    \end{flalign}
where $E_{\mathbf{k}}=k^{2}/2$ is the incident energy of the free electron, $\beta=1/k_{\mathrm{B}} T$, $\lambda_{T} = \sqrt{2\pi\beta}$ is the thermal deBroglie wavelength, and $n_{e}$ is the electron density. The first two terms inside the square brackets ($\left[...\right]$) correspond to broadening as a result of independent perturbations to the upper and lower states, respectively. The third term is an interference term that corrects for correlations between the perturbations to the upper and lower states. 

The $T$-matrix describes the electron-atom interactions. It is formally defined as 
\begin{eqnarray}
        \label{eq:Tmatrix}
        \hat{T}(E) = \sum_{n=0}^{\infty}\left[\hat{V}\frac{1}{E-\hat{H}_{0}}\right]^{n}\hat{V}
    \end{eqnarray}
where $\hat{H}_{0}$ is the non-interacting Hamiltonian and $\hat{V}$ is the interaction potential between the atom and the perturbing electron
    \begin{eqnarray}
    \label{eq:V}
        \hat{V} = -\frac{Z}{|\hat{\mathbf{r}}_{p}|}+\sum_{a}\frac{1}{|\hat{\mathbf{r}}_{a}-\hat{\mathbf{r}}_{p}|},
    \end{eqnarray}
where $\hat{\mathbf{r}}_{p}$ is the coordinate of the perturbing electron and $\hat{\mathbf{r}}_{a}$ is the coordinate of the atomic electron. The first term represents the free electron-nucleus interaction, and the second term represents the free electron-bound electron interaction.
A second order ($n=1$) $T$-matrix is used to calculate the upper and lower state terms, and a first order ($n=0$) $T$-matrix is used to calculate the interference term. This makes the three terms in Eq.~(\ref{eq:width2}) behave as $\mathcal{O}\left(V^{2}\right)$.

Importantly, the optical theorem~\cite{BarangerIII} can be used to further relate the upper and lower state broadening terms to total cross sections such that
\begin{eqnarray}
        \label{eq:opticaltheorem}
        \nonumber\mathrm{Im}\left\{\bra{\alpha\mathbf{k}}\hat{T}\left(E_{\alpha}+E_{\mathbf{k}}\right)\ket{\alpha\mathbf{k}}\right\} = -\frac{k}{2}\sum_{\alpha'}\sigma_{\alpha\rightarrow\alpha'}(\mathbf{k})\\
        \mathrm{Im}\left\{\bra{\beta\mathbf{k}}\hat{T}^{*}\left(E_{\beta}+E_{\mathbf{k}}\right)\ket{\beta\mathbf{k}}\right\} = \frac{k}{2}\sum_{\beta'}\sigma_{\beta\rightarrow\beta'}(\mathbf{k}).
    \end{eqnarray}
The total cross sections are written as sums over elastic ($\alpha=\alpha'$, $\beta=\beta'$) and inelastic ($\alpha\neq\alpha'$, $\beta\neq\beta'$) cross sections connecting the upper and lower states to other states in the system.
For a second order $T$-matrix, the cross sections (shown only for the upper state) become
\begin{eqnarray}
    \label{eq:crosssections}
    \sigma_{\alpha\rightarrow\alpha'}(\mathbf{k})\approx\frac{16\pi^{4} q}{k}\int d\Omega_{q}\left|\bra{\alpha\mathbf{k}}\hat{V}\ket{\alpha'\mathbf{q}}\right|^{2},
\end{eqnarray}
where $q = \sqrt{2\left(\mathrm{E}_{\alpha}+k^{2}/2-\mathrm{E}_{\alpha'}\right)}$ is the magnitude of the outgoing free electron wavenumber and $d\Omega_{q}$ is the solid angle element for the vector $\mathbf{q}$. It should be noted that this expression assumes that the radial free wavefunctions are normalized to a value of $\sqrt{(2/\pi)}/k$ in the asymptotic limit. More details on the calculation of the radial free wavefunctions are given in Sec.~\ref{subsec:AverageAtom}. 

The line width presented in this paper considers the following cross sections
    \begin{eqnarray}
        \label{eq:uppercross}
        \sum_{\alpha'}\sigma_{\alpha\rightarrow\alpha'}(\mathbf{k})\approx\sigma_{2p\rightarrow2s}(\mathbf{k})+\sigma_{2p\rightarrow2p}(\mathbf{k})\\
        \label{eq:lowercross}
        \sum_{\beta'}\sigma_{\beta\rightarrow\beta'}(\mathbf{k})\approx\sigma_{2s\rightarrow2s}(\mathbf{k}) + \sigma_{2s\rightarrow2p}(\mathbf{k}),
    \end{eqnarray}
where the $1$s$^{2}$ electrons that are common to both configurations have been omitted from the labels. Transitions to higher $n\ell$ values are omitted. Finally, the calculations here ignore exchange between the atomic and free electrons by assuming that the total wavefunction is a product state. Then, in order to calculate the cross sections and the resulting line width, it is necessary to determine the bound and free electron wavefunctions.
    
    \subsection{\label{subsec:AverageAtom}The Average-Atom Model}
    \begin{figure}[h!]\label{fig:bound}
        \includegraphics[width=0.5\textwidth]{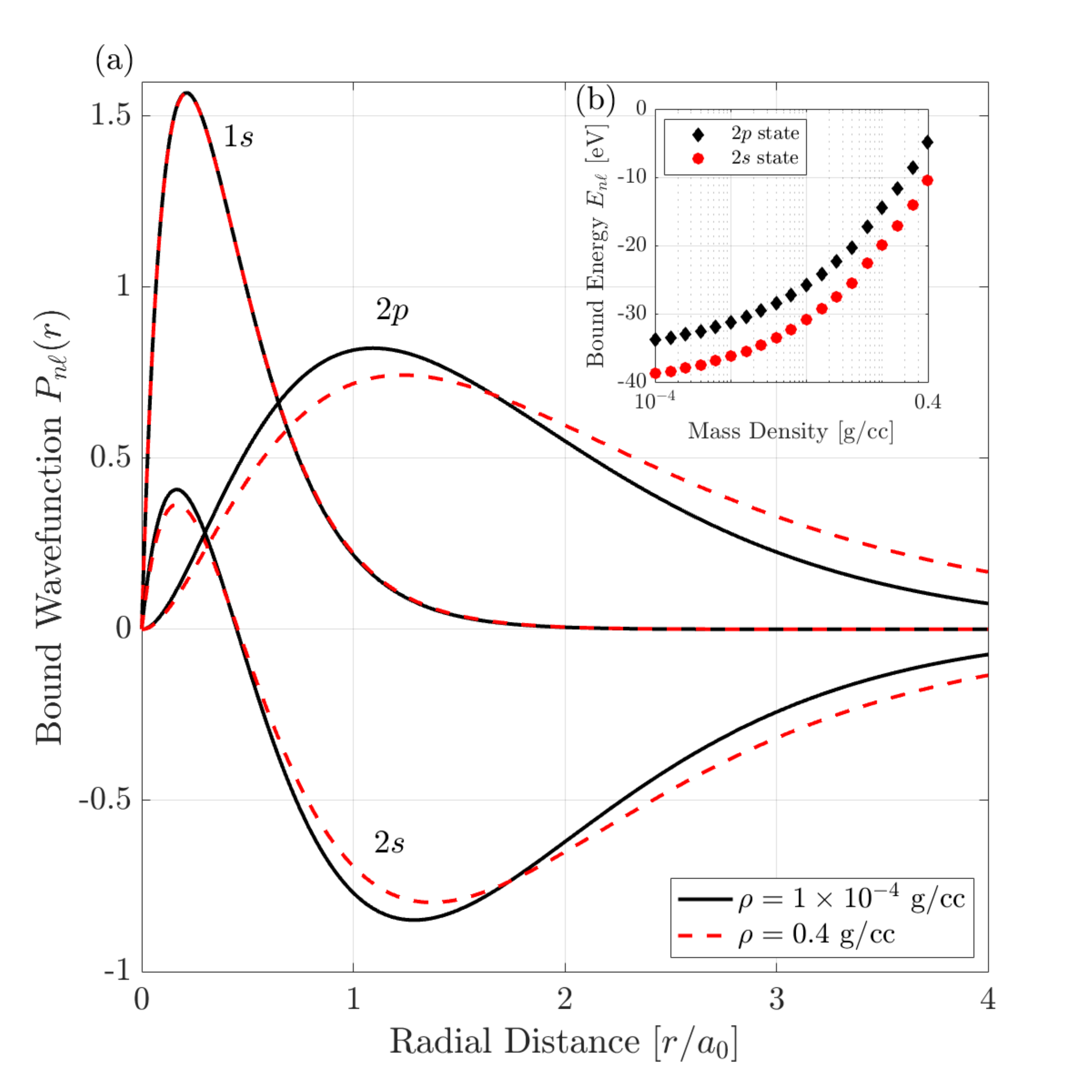}
        \caption{\label{fig:bound} (a) Radial bound wavefunctions computed in the AA model for the $1s$, $2s$, and $2p$ states. (b) Inset showing the increase in bound energies of the $2s$ and $2p$ states for increasing B~{\sc iii} mass density.}
        \end{figure}

The free and bound electronic wavefunctions were calculated from an AA model. Typical AA models ~\cite{wilson2006purgatorio,johnson2012thomson,starrett2014asimple} solve for the average configuration of bound and free charges and self-consistent potential around a central nucleus in a plasma of a given temperature $T$ and mass density $\rho$. 
    
First, the model makes an initial guess for the potential by interpolating between short-range Coulombic behavior ($-Z/r$) and zero at $r=R_\textrm{WS}$ where $R_\textrm{WS}=\left(3/(4\pi n_{i})\right)^{1/3}$ is the average inter-ionic spacing or the Wigner-Seitz radius. Then a solution to the radial Schrodinger equation
    \begin{eqnarray}
        \label{eq:RadSchrodinger}
        \frac{d^{2}P(r)}{dr^{2}}+2\left(E-V(r)-\frac{\ell(\ell+1)}{2r^{2}}\right)P(r)=0
    \end{eqnarray}
gives a set of radial bound ($P_{n\ell}$) and free ($P_{k\ell}$) wavefunctions. These wavefunctions are used to construct an electron density
    \begin{eqnarray}
        \label{eq:electrondensity}
        4\pi r^{2}n_{e}(r) = \sum_{a}f(E_{\mathbf{k}_{a}},\mu)g_{a}P_{a}^{2}(r),
    \end{eqnarray}
where $g_{a} = 2(2\ell_{a}+1)$ and $f(E_{\mathbf{k}_{a}},\mu)$ is a Fermi-Dirac distribution. The chemical potential is constrained by enforcing charge neutrality within the Wigner-Seitz radius such that
    \begin{eqnarray}
        \label{eq:muconstraint}
        Z = 4\pi\int_{0}^{R_\textrm{WS}}drr^{2}n_{e}(r).
    \end{eqnarray}
    Finally, the electron density is used to calculate a new potential
    \begin{eqnarray}
        \label{eq:AApotential}
        V(r) = -\frac{Z}{r}+\int_{r'<R_\textrm{WS}}d\mathbf{r}'\frac{n_{e}(r')}{|\mathbf{r}-\mathbf{r}'|}+V_{xc}(r).
    \end{eqnarray}
Here, the last term is the exchange-correlation potential $V_{xc}(r)=-(3n_{e}/\pi)^{1/3}+V_{c}(r)$, which in this model uses the local density approximation combined with a Hedin-Lundquist correlation potential~\cite{hedin1971explicit}. Equations~(\ref{eq:RadSchrodinger})-(\ref{eq:AApotential}) are solved self-consistently to give an electron density $n_{e}(r)$ and an AA potential $V(r)$. In order to accurately represent the `frozen core' of the boron atom, the model used here sets $f\left(E_{\mathbf{k}_{a}},\mu\right) = 1$ for the two $1s$ electrons and $f\left(E_{\mathbf{k}_{a}},\mu\right) = 0$ for all other bound states. This ensures the bound contribution to the electron density and the AA potential is solely due to the `frozen core' of $1s$ electrons, isolating the effect of density (rather than changes in the average configuration) and removing self-screening from the orbitals of the active bound electron.  

Figure~\ref{fig:bound} gives binding energies and bound wavefunctions
for a range of mass densities $\rho = 10^{-4}-0.4$ g/cc. 
Increasing the density brings the bound state wavefunctions for the $2s$ and $2p$ electrons further away from the nucleus, and increases their energies. The $1s$ electrons are much more tightly bound, and their wavefunctions do not change appreciably throughout the density range examined here. Their 
binding energy varies from $-227$ to $-191$~eV over the density range considered here~$\rho=10^{-4}-0.4$ g/cc. 

    \begin{figure}[h!]\label{fig:DOS}
        \includegraphics[width=0.5\textwidth]{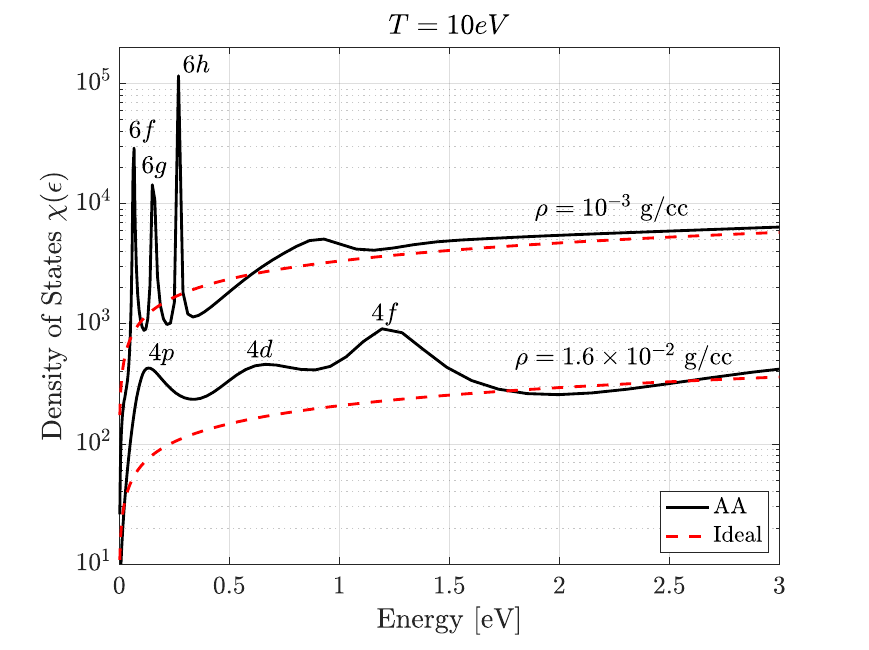}
        \caption{\label{fig:DOS} Continuum DOS of B~{\sc iii} at $T = 10$ eV as a function of free electron energy. The solid-black line corresponds to the solution from Eq.~(\ref{eq:continuumDOS}) with free electron wavefunctions calculated in the AA potential. The dashed-red line is the ideal continuum DOS $\chi_{i}(E_{\mathbf{k}})=\sqrt{2E_{\mathbf{k}}}/\left(n_{i}\pi^{2}\right)$.}
        \end{figure}

Importantly, the AA potential models screening from the surrounding plasma particles and pressure-ionized bound states. Screening is captured by the boundary condition $V(r=R_\textrm{WS})=0$ and the contribution of the free electron density in the second term of Eq.~(\ref{eq:AApotential}). As screening increases and the Wigner-Seitz radius decreases with density, the binding energies of all $n\ell$ orbitals move towards the continuum limit. When an orbital can no longer be supported by the self-consistent potential, it becomes pressure ionized and forms a continuous ``resonance'' in the continuum density of states (DOS).
The continuum DOS can be calculated from the free wavefunctions
    \begin{eqnarray}
        \label{eq:continuumDOS}
        \chi(E_{\mathbf{k}}) = \sqrt{2E_{\mathbf{k}}}\sum_{\ell}2(2\ell+1)\int_{0}^{R_\textrm{WS}} drP^{2}_{k\ell}(r). 
    \end{eqnarray}
Figure~\ref{fig:DOS} shows that the pressure-ionized bound states modify the continuum DOS at low energies. At high energies, the continuum DOS approaches the ideal value $\chi_{i}(E_{\mathbf{k}}) = \sqrt{2E_{\mathbf{k}}}/(n_{i}\pi^{2})$. As the density increases and the AA potential is restricted to a smaller radius, lower $n\ell$ orbitals
will appear in the continuum. For example, only the $n=6$ states appear in the continuum at $\rho=10^{-3}$~g/cc. However, at a density of $\rho=1.6\times10^{-2}$~g/cc, the $n=4$ states emerge.  Furthermore, increasing density modifies the continuum DOS at higher energies. For example, at a density of $\rho=10^{-3}$~g/cc the resonances in the continuum DOS only extend out to $\sim1$~eV, while in the case of $\rho=1.6\times10^{-2}$~g/cc the resonances extend further out to $\sim2$~eV. These resonances in the continuum DOS are encoded in the free electron wavefunctions themselves and will impact the calculation of the cross sections and the line width.

We note that while the AA model includes self-consistent screening that modifies both bound and free electron wavefunctions, an additional screening effect may be invoked in the interaction potential given by Eq.~(\ref{eq:V}) that is used to evaluate matrix elements of the form in Eq.~(\ref{eq:crosssections}). This will be considered in Sec.~\ref{subsec:LineWidth}.

\subsection{\label{subsec:BetheFormula}Bethe Formula from the Impact Approximation}

Although the impact approximation in Sec.~\ref{sec:Theory} is already derived from a strict set of physical assumptions, many of the atomic codes used to predict detailed emission and opacity spectra~\cite{CHUNG20053,DCA} in complex, multi-electron ions do not have direct access to wavefunctions and instead rely on collisional cross sections and rates to model electron-collision broadening. To connect the present work with these codes, we invoke additional assumptions. One widely used approximation is the Bethe formula, which writes the inelastic electron-collision cross section in terms of an oscillator strength and a free-free Gaunt factor~\cite{bethe1930theory}. The Bethe formula can be derived from the impact approximation beginning with the inelastic electron-collision cross sections appearing in Eqs.~(\ref{eq:uppercross}) and~(\ref{eq:lowercross}). For optically allowed transitions ($|\ell_{\alpha}-\ell_{\alpha'}| = 1$), the excitation cross section computed from Eq.~(\ref{eq:crosssections}) is
\begin{flalign}
    \label{eq:crossinelastic1}
    \nonumber&\sigma_{\alpha\rightarrow\alpha'}(k) = \frac{1}{4\pi}\int d\Omega_{k}\sigma_{\alpha\rightarrow\alpha'}(\mathbf{k})\\ &= \sum_{\ell}\frac{4\pi^{3} q\left(\ell+1\right)\ell_{>}}{3k(2\ell_{\alpha}+1)}\left[\mathcal{R}_{\substack{\alpha,\alpha'\\\ell+1,\ell}}^{2}+\mathcal{R}_{\substack{\alpha,\alpha'\\\ell,\ell+1}}^{2}\right]
\end{flalign}
where $\ell_{>}$ is the greater of $\ell_{\alpha}$ and $\ell_{\alpha'}$ and
\begin{flalign}
    \label{eq:radialintegrals}
    \nonumber\mathcal{R}_{\substack{\alpha,\alpha'\\\ell,\ell'}} = \int_{0}^{\infty}\int_{0}^{\infty}dr_{r}dr_{p}&P_{n_{\alpha}\ell_{\alpha}}(r_{r})P_{n_{\alpha'}\ell_{\alpha'}}(r_{r})\\\times &P_{k\ell}(r_{p})P_{q\ell'}(r_{p})\left\{\frac{r_{<}}{r_{>}^{2}}\right\}
\end{flalign}
is a double integral involving the pre- and post-collision bound ($P_{n_{\alpha}\ell_{\alpha}}$,$P_{n_{\alpha'}\ell_{\alpha'}}$) and free ($P_{k\ell}$,$P_{q\ell'}$) radial wavefunctions. Here, $r_{r}$ represents the radial position coordinate of the radiating electron and $r_{<}$ $(r_{>})$ represents the lesser (greater) of $r_{r}$ and $r_{p}$ respectively. The final term in braces $\left\{...\right\}$ makes the two integrals inseparable. The crucial approximation in deriving the Bethe formula is to assume that the free electron is always further away from the nucleus than the bound electron. This amounts to saying $r_{<}\approx r_{r}$ and $r_{>}\approx r_{p}$, so that
\begin{eqnarray}
    \label{eq:Bethe}
    \sigma_{\alpha\rightarrow\alpha'}(k) \approx \frac{4\pi^{2}f_{\alpha,\alpha'}g_\textrm{ff}\left(k,q\right)}{\sqrt{3} E_{\alpha,\alpha'}k^{2}},
\end{eqnarray}
where 
\begin{eqnarray}
    \label{eq:osc}
    f_{\alpha,\alpha'} = \frac{2E_{\alpha,\alpha'}\ell_{>}}{3(2\ell_{\alpha}+1)}\left[\int_{0}^{\infty}dr_{r}P_{n_{\alpha}\ell_{\alpha}}r_{r}P_{n_{\alpha'}\ell_{\alpha'}}\right]^{2}
\end{eqnarray}
is the oscillator strength, $E_{\alpha,\alpha'} = |E_{\alpha}-E_{\alpha'}|$ is the energy of the inelastic transition, and
\begin{eqnarray}
    \label{eq:gff}
    g_\textrm{ff}\left(k,q\right) = \frac{\sqrt{3}\pi kq}{2}\sum_{\ell}(\ell+1)\left[\mathcal{F}^{2}_{\substack{k,q\\\ell+1,\ell}}+\mathcal{F}^{2}_{\substack{k,q\\\ell,\ell+1}}\right]
\end{eqnarray}
is the hydrogenic ($Z = 1$) energy-dependent free-free Gaunt factor $g_\textrm{ff}(k,\omega)$ evaluated at a frequency $\omega = \left(q^{2}-k^{2}\right)/2$. The Gaunt factor is written in terms of the radial integrals
\begin{eqnarray}
    \label{eq:gff_integral}
    \mathcal{F}_{\substack{k,q\\\ell,\ell'}} = \int_{0}^{\infty}dr_{p}P_{k\ell}\frac{1}{r_{p}^{2}}P_{q\ell'}.
\end{eqnarray}
 Equation~(\ref{eq:Bethe}) represents the Bethe formula. It is important to note that, when evaluated using free wavefunctions from the AA model, the real part of the free-free dynamic conductivity ($\sigma(\omega)$) of a plasma is given in the Kubo-Greenwood approximation by
\begin{eqnarray}
    \label{eq:KuboGreenwood}
    \mathrm{Re}\left\{\sigma(\omega)\right\} = \frac{4\pi^{2}n_{e}n_{i}}{3\sqrt{3}\omega^{3}}\left(1-e^{-\beta\omega}\right)\langle\frac{1}{k}g_{\textrm{ff,AA}}(k,\omega)\rangle,
\end{eqnarray}
where $\langle...\rangle$ is the thermal average of the energy-dependent free-free Gaunt factor over incident electron energies. Here, $g_{\textrm{ff,AA}}$ is almost identical to Eq.~(\ref{eq:gff}), except that the Coulomb force ($1/r_{p}^{2}$) is replaced by the force due to an effective electron-atom interaction ($-dV(r_{p})/dr_{p}$) where $V(r_{p})$ is the AA potential in Eq.~(\ref{eq:AApotential}). Thus, in this approximation, the same procedures that are used to calculate dynamic conductivity in dense plasmas can be used to approximate line widths as well. 

\section{\label{sec:Discussion}Results}

The effects of screening and pressure ionization are included in the free wavefunctions calculated from the AA potential. Sections~\ref{subsec:CrossSections} and~\ref{subsec:LineWidth} examine the impact of this physics on the electron-collision cross sections and line width by comparison with identical calculations that use free wavefunctions calculated in a bare Coulomb potential. Section~\ref{subsec:LineWidth} also considers a screened interaction potential as a substitute for Eq.~(\ref{eq:V}) to show that whether or not there is screening in the wavefunction, screening in the matrix element itself can have a significant effect on the line width. Section~\ref{subsec:BetheFormulaComparison} compares the AA calculations with inelastic cross sections calculated from the Bethe formula in Eq.~(\ref{eq:Bethe}). 

\subsection{\label{subsec:CrossSections}Cross Sections from the Average Atom Model}

\begin{figure}[htbp!]\label{fig:cross_sections}
\includegraphics[width=0.5\textwidth]{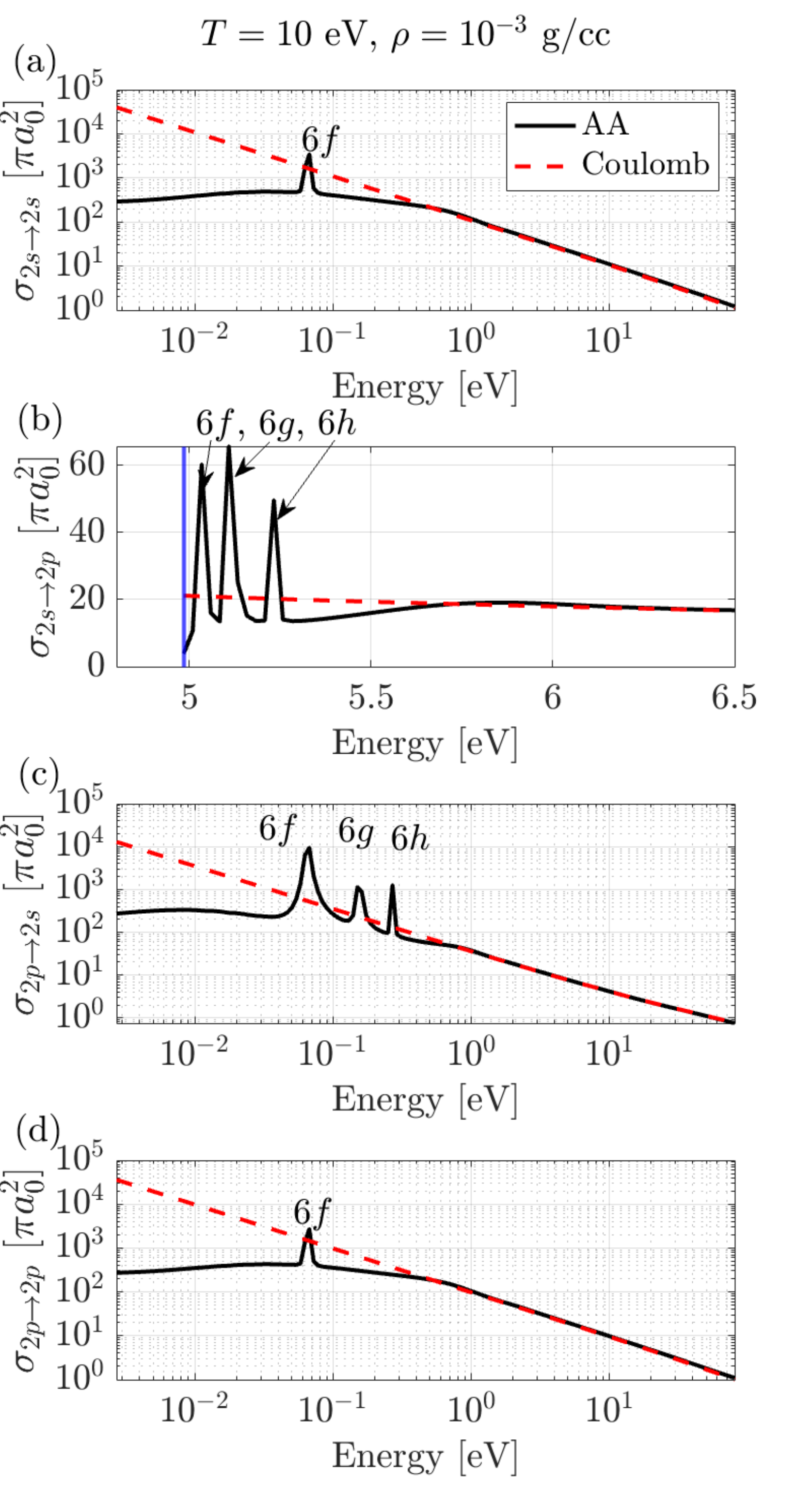}
\caption{\label{fig:cross_sections} Cross sections for the electron-collision cross sections appearing in Eqs.~(\ref{eq:uppercross}) and~(\ref{eq:lowercross}) as a function of the incoming free electron energy for a mass density of $\rho = 10^{-3}$g/cc and temperature $T = 10$eV. The two curves compare calculations based on free-electron wavefunctions calculated in the AA (solid black) and Coulomb (dashed red) potentials. The vertical blue line in (b) represents the threshold energy needed for the inelastic ($\sigma_{2s\rightarrow2p}$) cross section to be nonzero. The cross sections calculated using the AA potential show a reduced value at low energies and resonances that are due to self-consistent screening and the presence of pressure-ionized bound states.}
\end{figure}

\begin{figure}[htbp!]\label{fig:2s2p}
\includegraphics[width=0.5\textwidth]{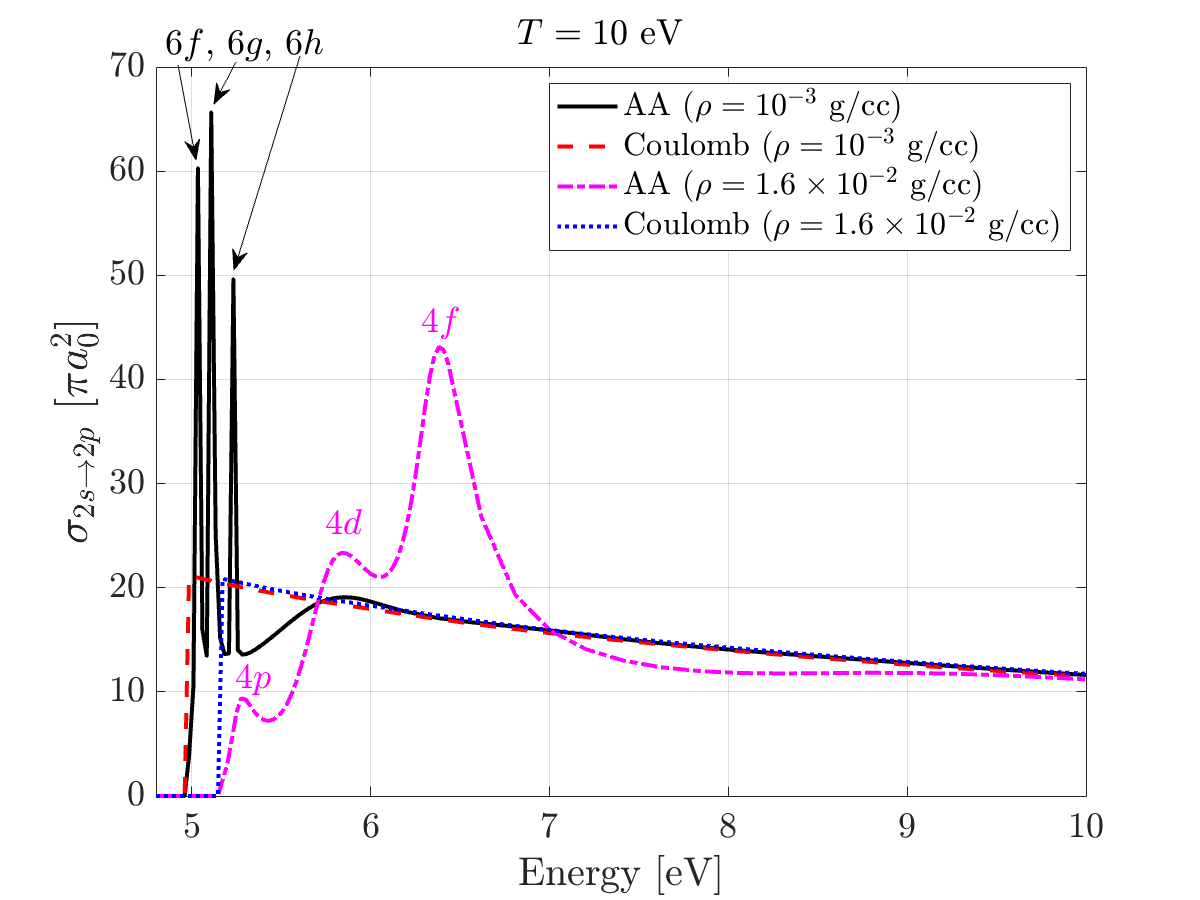}
\caption{\label{fig:2s2p} Inelastic cross section for the $2s\rightarrow2p$ transition at two different densities using AA and Coulomb free wavefunctions. Modifications due to screening and pressure ionization from the AA free wavefunctions move to higher energy as the density increases.}
\end{figure}

\begin{figure*}[htbp!]\label{fig:linewidth}
\includegraphics[width=1\textwidth]{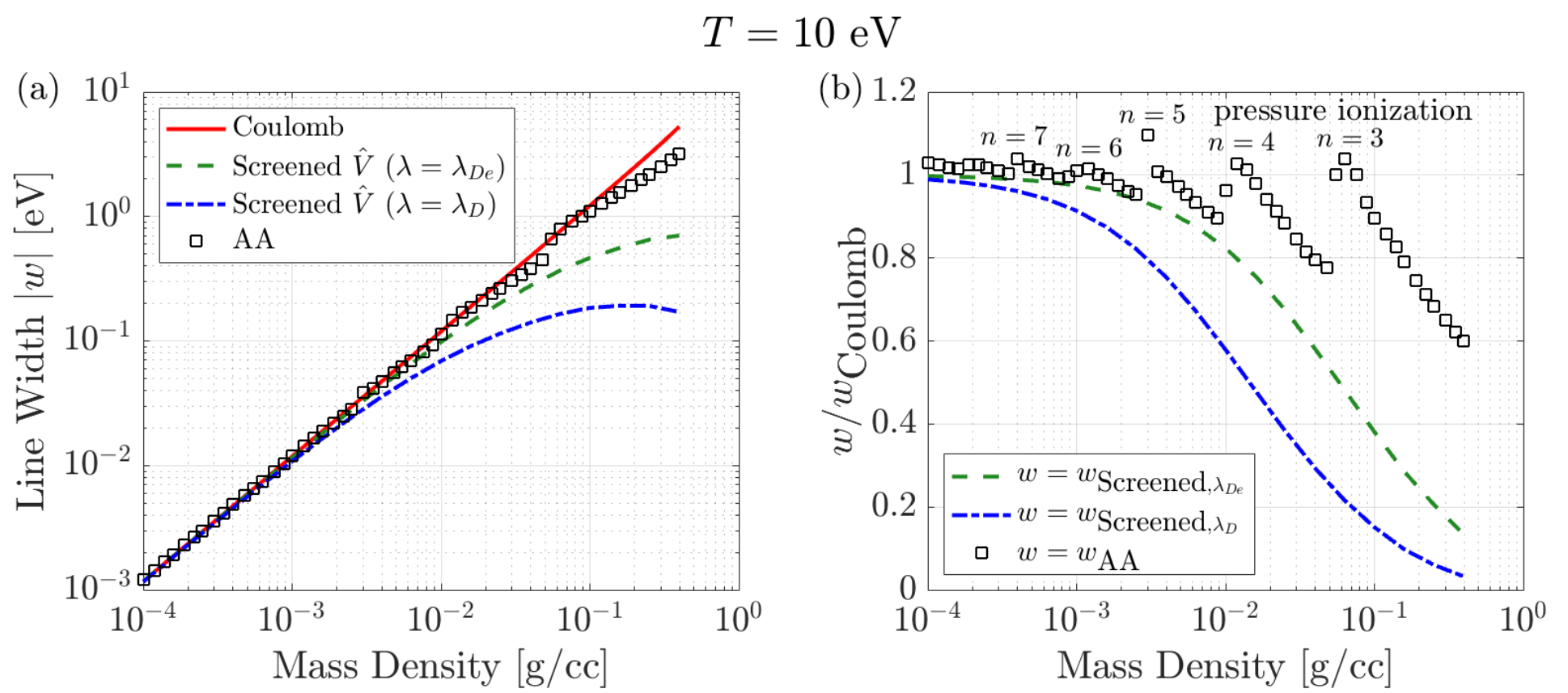}
\caption{\label{fig:linewidth} (a) Line widths calculated using AA (open squares) and Coulomb (solid red line) free wavefunctions differ at higher density. These two calculations use the Coulomb interaction potential given by Eq.~(\ref{eq:V}) inside the matrix elements. The dashed lines show the line width calculated using the Coulomb free wavefunctions and the screened interaction potential in Eq.~(\ref{eq:Vscr}). (b) Relative line width between AA and Coulomb free wavefunctions shows a general decrease due to screening and sharp increases due to pressure ionization of bound states. The principal quantum number of the states corresponding to the pressure ionization peaks are labeled accordingly.}
\end{figure*}

The free wavefunctions calculated in the AA potential modify both the elastic and inelastic cross sections compared to the Coulomb result (see Fig.~\ref{fig:cross_sections}). First, the inclusion of screening due to surrounding plasma particles in the AA potential reduces the strength of the potential further away from the nucleus. This primarily impacts low energy electron states which are located further away from the nucleus than high energy electron states. This is corroborated by Figs.~\ref{fig:cross_sections}(a), \ref{fig:cross_sections}(c), and \ref{fig:cross_sections}(d), which all show reduced cross sections at low incoming electron energy. In the case of the inelastic $\sigma_{2s\rightarrow2p}$ cross section in Fig.~\ref{fig:cross_sections}(b) there is a threshold energy equal to the energy difference of the 2$s$ and 2$p$ states, so the screening reduction in the cross section occurs at low energies relative to the threshold. At high energies, the AA and Coulomb cross sections converge to the same value.

Pressure-ionized bound states also influence the cross sections by producing resonances at their associated energies. For example, at a mass density $\rho = 10^{-3}$ g/cc the continuum DOS from the AA model showed pressure-ionized bound states corresponding to the $6f$, $6g$, and $6h$ orbitals (see Fig.~\ref{fig:DOS}). The inelastic $\sigma_{2p\rightarrow2s}$ cross section shows resonances at energies corresponding to these states. In the case of the inelastic $\sigma_{2s\rightarrow2p}$ cross section, the resonances show up at energies relative to the threshold energy. For the elastic cross sections, only a resonance corresponding to the $6f$ state appears.

Finally, increasing the density modifies the AA cross sections at higher energies. This is shown in Fig.~\ref{fig:2s2p}, where the reduction of the cross section near the threshold due to the AA free wavefunctions is enhanced and persists to a higher energy as the density increases. The resonances also modify the cross section at higher energy as the density increases. Because Eq.~(\ref{eq:width2}) involves integration of the cross sections over a Maxwellian energy distribution of free electrons at $T = 10$ eV, the appearance of these features at higher energy significantly influence the spectral line width. 


\subsection{\label{subsec:LineWidth}Effect on the Line Width} 
The line width was calculated from the upper, lower, and interference terms using Eq.~(\ref{eq:width2}). 
The free electron density in this formula was modeled using $n_{e} = \Bar{Z}n_{i}$ with a constant average ionization of $\Bar{Z} = 3$, consistent with the `frozen core' picture of the B~{\sc iii} atom. Line widths calculated using both the AA and Coulomb free wavefunctions are predicted to increase with increasing density (see Fig.~\ref{fig:linewidth}(a)). This is expected as a larger free electron population should increase collisional broadening. The AA calculation begins to differ from the purely Coulomb one at high density ($\rho \gtrsim 10^{-3}$~g/cc). The screening and pressure ionization that drive these differences are best highlighted by comparing the relative line width between the AA and Coulomb free wave function cases (see Fig.~\ref{fig:linewidth}(b)). 

First, the resonances due to pressure-ionized bound states in the AA model increase the line width relative to the Coulomb case when a set of bound states corresponding to a certain principal quantum number $n$ are pressure ionized. For example, the peak in the relative line width around $\rho\sim10^{-2}$ g/cc is a result of the pressure ionization of the $n = 4$ states. With a further increase in density, the $n = 4$ states blend into the continuum and the relative line width continues the general trend of decreasing (see Fig.~\ref{fig:DOSevo}). Second, the relative line width generally decreases as the density increases. This is because at higher densities the screening modifications that lower the cross sections are pushed to higher energies. Overall, the screening and pressure ionization introduced by the AA free wavefunctions have a considerable impact on the relative line width, creating variations of around $40\%$ at the highest densities explored in this paper. 

The finding that screening reduces the line width is also qualitatively consistent with previous work~\cite{griem1959Stark} and supported by the well-known method of screening the atom-electron interaction potential in Eq.~(\ref{eq:V})~\cite{gomez2022intro}. For a Debye screened interaction, the new potential is
\begin{eqnarray}
    \label{eq:Vscr}
    \hat{V} = -\frac{Ze^{-|\hat{\mathbf{r}}_{p}|/\lambda}}{|\hat{\mathbf{r}}_{p}|}+\sum_{a}\frac{e^{-|\hat{\mathbf{r}}_{a}-\hat{\mathbf{r}}_{p}|/\lambda}}{|\hat{\mathbf{r}}_{a}-\hat{\mathbf{r}}_{p}|}
\end{eqnarray}
where $\lambda$ can be the total Debye length $\lambda_\textrm{D} = \sqrt{T/\left(4\pi n_{e}(\bar{Z}+1)\right)}$ or the electron Debye length $\lambda_{\textrm{D}e} = \sqrt{T/\left(4\pi n_{e}\right)}$. Both choices are shown in Fig.~{\ref{fig:linewidth}} and significantly reduce the line width at high density. It is noted that the calculations with the screened interaction potential still use Coulomb free wavefunctions, and thus do not capture any of the structure due to the pressure ionization. We also note that the AA widths do not include the screening term in the interaction potential.

\begin{figure}[htbp!]\label{fig:DOSevo}
\includegraphics[width=0.5\textwidth]{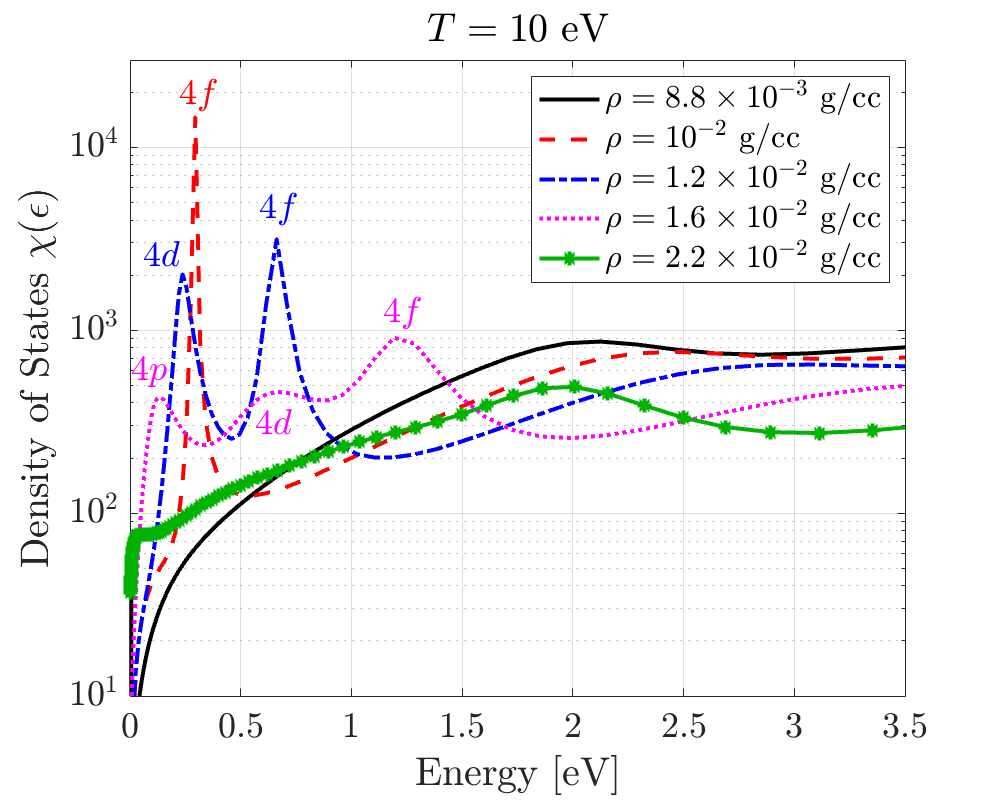}
\caption{\label{fig:DOSevo} The continuum DOS for mass density range of $\rho=8.8\times10^{-3}-2.2\times10^{-2}$~g/cc. The pressure ionization of the $n=4$ states drives the increase in relative line width around $\rho\sim10^{2}$~g/cc in Fig.~\ref{fig:linewidth}(b).}
\end{figure}

\subsection{\label{subsec:BetheFormulaComparison}Comparison with the Bethe Formula}
In order to evaluate the Bethe formula in Eq.~(\ref{eq:Bethe}), the oscillator strengths are calculated using the bound wavefunctions from the AA model. Then, the Gaunt factor is calculated using both the Coulomb and AA free wavefunctions. The Bethe formula generally overpredicts the cross sections compared to the impact approximation calculation (see Figs.~\ref{fig:bethe}(a) and~\ref{fig:bethe}(b)). The physical reasoning for this result is that the approximation in the derivation of Eq.~(\ref{eq:Bethe}) leads to a larger electron-electron interaction potential at close distances. The radial integral appearing in Eq.~(\ref{eq:radialintegrals}) can be written as
\begin{eqnarray}
    \label{eq:radialintegralsVeff}
    R_{\substack{\alpha,\alpha'\\\ell,\ell'}} = \int_{0}^{\infty}dr_{p}P_{k\ell}(r_{p})V_\textrm{eff}(r_{p})P_{q\ell'}(r_{p})
\end{eqnarray}
where 
\begin{flalign}
    \label{eq:Veff}
    \nonumber V_\textrm{eff}(r_{p}) = \frac{1}{r_{p}^{2}}\int_{0}^{r_{p}}dr_{r}P_{n_{\alpha}\ell_{\alpha}}(r_{r})r_{r}P_{n_{\alpha'}\ell_{\alpha'}}(r_{r})\\
    +r_{p}\int_{r_{p}}^{\infty}dr_{r}P_{n_{\alpha}\ell_{\alpha}}(r_{r})\frac{1}{r_{r}^{2}}P_{n_{\alpha'}\ell_{\alpha'}}(r_{r})
\end{flalign}
is the effective electron-electron interaction potential. To derive the Bethe formula, the effective potential is written as
\begin{eqnarray}
    \label{eq:Veffapprox}
    V_\textrm{eff,Bethe}(r_{p})\approx\frac{1}{r_{p}^{2}}\int_{0}^{\infty}dr_{r}P_{n_{\alpha}\ell_{\alpha}}(r_{r})r_{r}P_{n_{\alpha'}\ell_{\alpha'}}(r_{r}).
\end{eqnarray}
Figure~\ref{fig:bethe}(c) shows that the consequence of this approximation is to increase the effective potential at short distances. 

The $\sigma_{2s\rightarrow2p}$ inelastic cross section can also be compared with a standard analytic result in the literature from Mewe~\cite{mewe1972interpolation} that approximates the Gaunt factor for the excitation process within a given $n$ shell as 
\begin{eqnarray}
    \label{eq:Mewe}
    g_\textrm{ff}(k,q)\approx0.6+\sqrt{3}/(2\pi)\ln{\left(k^{2}/2/E_{2s,2p}\right)},
\end{eqnarray}
where in this paper $E_{2s,2p}$ is evaluated using the energy levels from the AA model. This simple formula shows relatively good agreement with the full impact approximation results that use a Coulomb potential. However, it cannot capture the reduction in the cross-section at low energies and the resonances due to the AA free wavefunctions. 

In the end, the impact of the Bethe approximation on the line width is significant. The relative line widths shown in Figure~\ref{fig:bethe}(d) were obtained using the same procedure for the impact approximation outlined in Sec.~{\ref{subsec:LineBroadening}}, but using the Bethe approximation to calculate both the $\sigma_{2s\rightarrow2p}$ and the $\sigma_{2p\rightarrow2s}$ inelastic cross sections. Even at low density, the Bethe formula increases the relative line width by $\sim10\%$ and $\sim40\%$ for AA and Coulomb free wavefunctions, respectively.


\begin{figure}[htbp!]\label{fig:bethe}
\includegraphics[width=0.5\textwidth]{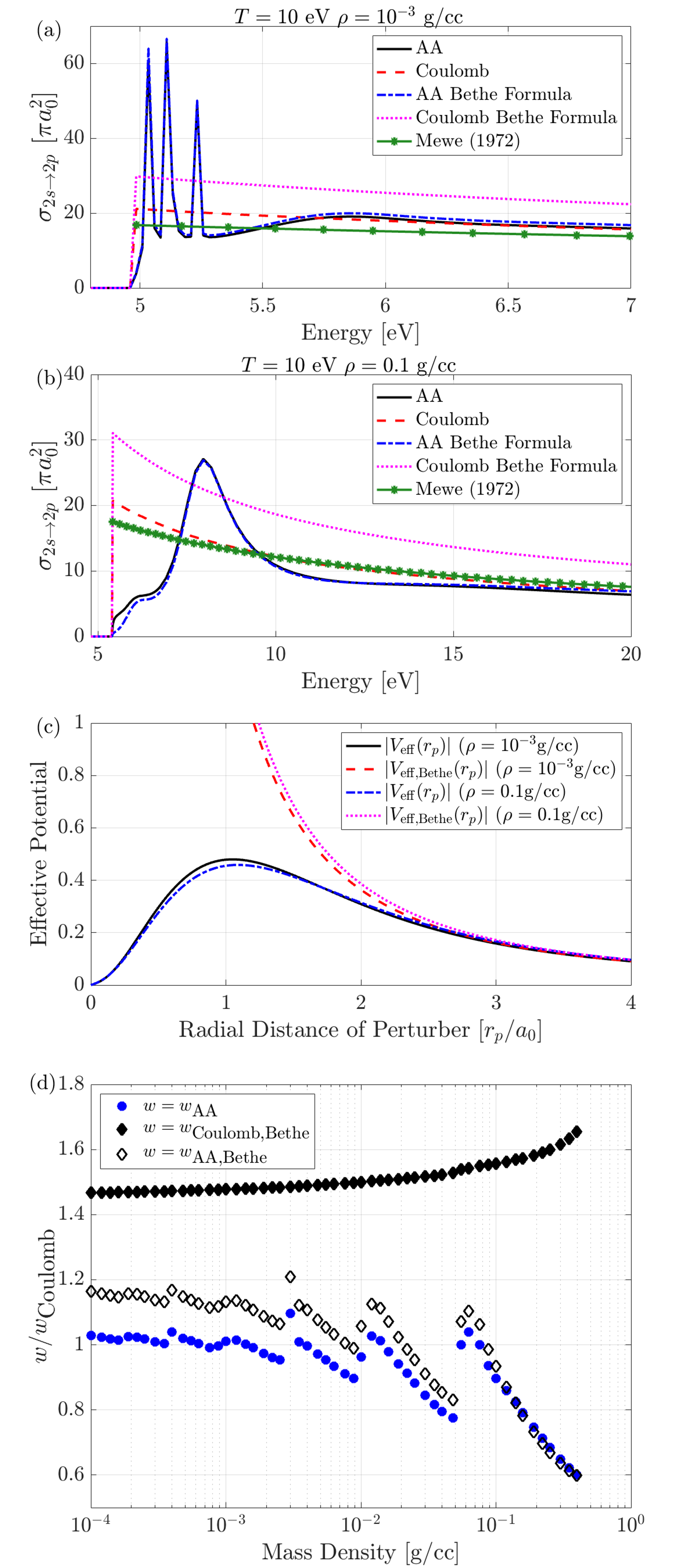}
\caption{\label{fig:bethe} (a,b) Calculations of the $\sigma_{2s\rightarrow2p}$ cross section using Eq.~(\ref{eq:crosssections}) (solid black and dashed red lines) compared with the Bethe formula (dash-dotted blue and dotted magenta lines). AA and Coulomb refer to the potential used to calculate the free electron wavefunctions. (c) The overprediction of the cross sections by the Bethe formula is a result of the approximation to the effective electron-electron interaction potential at close distances. (d) The line width calculated using the Bethe formula is higher than the calculation using the impact approximation. Here, the Bethe formula is used for both the $\sigma_{2s\rightarrow2p}$ and $\sigma_{2p\rightarrow2s}$ inelastic cross sections.}
\end{figure}

\section{\label{sec:conclusion}Conclusion}

Accurate calculation of spectral line widths in plasmas involves a complicated $N$ body problem, where the effect of plasma electrons and ions on the radiative process must be taken into account. This work used an AA model to calculate self-consistent free-electron wavefunctions and study their physical influence on the line width. Through the free electron wavefunctions, the AA potential introduces screening and pressure-ionized bound states. Compared with identical calculations using free wavefunctions from a Coulomb potential, the physics of screening is shown to generally reduce the relative line width as the plasma density increases. This observation is consistent with the common method of screening the actual interaction potential in the matrix element. However, it is found that screening the interaction potential in the matrix element reduces the line width much more than screening the free wavefunctions. Meanwhile, pressure ionization causes an increase in the relative line width as bound states appear in the continuum.

Although the AA model is used here to emphasize the importance of describing screening and pressure ionization in spectral line shape theory, future work is needed to incorporate the AA framework into a more rigorous line shape calculation. First, although this paper shows that both the potential in the matrix element and the potential used to find the bound and free wavefunctions influence the line width, there is a need for consistency in how each is computed. Furthermore, the calculation here uses lower-order $T$-matrices and ignores exchange between electrons, both of which are approximations that are possible to remove~\cite{gomez2021allorder}. Future work should remove these approximations so that an AA informed calculation like this may undergo further scrutiny. In a very dense plasma, even the assumption that electron-atom collisions are statistically independent becomes invalid, calling the impact approximation itself into question. Finally, there remain question about the lifetimes of continuum resonances~\cite{Piron2026}. Still, this work suggests that the physics of free electrons, when treated using an AA model, may have a measurable impact on spectral line widths in high density plasmas.

\begin{acknowledgments}
This work is funded by the NNSA Stockpile Stewardship Academic Alliances under Grant No. DE-NA0004100 and the DOE NNSA Stockpile Stewardship Graduate Fellowship through cooperative agreement DE-NA0003960. Additionally, this research was supported in part through computational resources and services provided by Advanced Research Computing (ARC), a division of Information and Technology Services (ITS) at the University of Michigan, Ann Arbor. T. A. G. acknowledges support from the George Ellery Hale Post-Doctoral Fellowship at the University of Colorado. Sandia National Laboratories is a multi-mission laboratory managed and operated by National Technology \& Engineering Solutions of Sandia, LLC (NTESS), a wholly owned subsidiary of Honeywell International Inc., for the U.S. Department of Energy’s National Nuclear Security Administration (DOE/NNSA) under contract DE-NA0003525. This written work is authored by an employee of NTESS. The employee, not NTESS, owns the right, title and interest in and to the written work and is responsible for its contents. Any subjective views or opinions that might be expressed in the written work do not necessarily represent the views of the U.S. Government. The publisher acknowledges that the U.S. Government retains a non-exclusive, paid-up, irrevocable, world-wide license to publish or reproduce the published form of this written work or allow others to do so, for U.S. Government purposes. The DOE will provide public access to results of federally sponsored research in accordance with the DOE Public Access Plan.
\end{acknowledgments}

\section*{Data Availability Statement}
The data that support the findings of this study are available from the corresponding author upon reasonable request.

\bibliography{aipsamp}

\end{document}